 \documentclass[final,5p,times,twocolumn,authoryear]{elsarticle}

\usepackage{amssymb}
\usepackage{lipsum}
\global\arraycolsep=2pt
\usepackage{amsmath}
\usepackage{graphicx}
\usepackage{textcomp}
\usepackage{calrsfs}
\usepackage{caption}
\usepackage{subcaption}
\usepackage{bm}
\usepackage[utf8]{inputenc}
\usepackage{textgreek}
\usepackage{color}
\usepackage{multirow}
\usepackage{array}
\usepackage{orcidlink}
\newcommand{\ben}{\begin{equation*}}
\newcommand{\een}{\end{equation*}}
\newcommand{\bean}{\begin{eqnarray*}}
\newcommand{\eean}{\end{eqnarray*}}

\newcommand{\be}{\begin{equation}} 
\newcommand{\ee}{\end{equation}}
\newcommand{\bea}{\begin{eqnarray}}
\newcommand{\eea}{\end{eqnarray}}
\newcommand{\ba}{\begin{aligned}}
\newcommand{\ea}{\end{aligned}}

\usepackage{hyperref}
\hypersetup{
     colorlinks = true,
     citecolor  = magenta,  
     linkcolor  = magenta 
}

\journal{Annals of Physics}

\begin{document}

\begin{frontmatter}



\title{Casimir Forces Across Magnetic Plasmas at Nuclear Separations  }


\author[First]{S. K. Panja \,\orcidlink{0000-0001-9258-9825}\,}
\ead{sumanpanja19@gmail.com}
\affiliation[First]{organization={Centre of Excellence ENSEMBLE3},
            addressline={Sp. z o. o., Wolczynska Str. 133}, 
            city={Warsaw},
            postcode={01-919}, 
            country={Poland}}
\author[First]{L. Inacio\,\orcidlink{0000-0001-8971-0591}}
\author[Second]{S. Pal\,\orcidlink{0009-0003-6356-3409}\,}
 \affiliation[Second]{organization={Dipartimento di Fisica e Chimica - Emilio Segrè, Università degli Studi di Palermo},
addressline={Via Archirafi 36}, 
city={Palermo}, 
postcode={90123},
country={Italy}
}
\author[First,Forth]{M. Bostr\"om\,\orcidlink{0000-0001-5111-4049}}
\ead{mathias.bostrom@ensemble3.eu}
            
\affiliation[Forth]{organization={Chemical and Biological Systems Simulation Lab, Centre of New Technologies, University of Warsaw},
            addressline={Banacha 2C}, 
            city={Warsaw},
            postcode={02-097}, 
            country={Poland}}

\begin{abstract}
A theory and numerical findings are presented on the magnetic Casimir interaction that arises from vacuum fluctuations of the quantized field and its effects at the nuclear scale. We investigate how the zero-temperature Casimir effect at nuclear scales can generate the black-body temperatures required to induce a magnetic electron-positron plasma. The magnetic permeability of the plasma and any magnetic fields present influence the screened Casimir-Yukawa potentials between perfect conducting surfaces. We discuss implications for the magnetic Casimir-Yukawa potential, its screening length, and a magnetic permeability-dependent quantity that resembles the meson mass. 
\end{abstract}




\end{frontmatter}

\section{Introduction}
\label{introduction}

The Casimir effect arises due to vacuum fluctuations of the quantized field. These vacuum fluctuations are restricted when two uncharged conducting plates are brought close together, with a minimal separation distance. This results in a pressure difference between the regions inside and outside the plates. This pressure difference results in an attractive force between the parallel plates. This effect, named after the Dutch physicist \,\cite{CH-0-HBG}, was first predicted in 1948. Casimir proposed a simplified model where molecules are replaced by uncharged, perfectly conducting plates and calculated the energy of the system arising purely due to the vacuum fluctuations of electromagnetic field. He obtained the energy and force per unit area between the plates as\,\citep{CH-0-HBG}
\begin{equation}
\frac{E_{C}}{A}=-\frac{\pi^{2}}{720 \,L^{3}}\hbar c,~~~~ \frac{F_{C}}{A}=-\frac{\pi^{2}}{240 \,L^{4}}\hbar c, \label{Cas-Nucl-1}
\end{equation}
respectively. Here, $\hbar$ is the reduced Planck constant, $c$ is the velocity of light, and $L$ is the separation distance between the parallel plates. \cite{lifshitz} used the regularization technique to eliminate the divergence in zero-point energy and successfully reproduced Casimir's original result. Moreover, the Lifshitz's formula naturally incorporates the effects of finite temperature and material properties, such as the dielectric response of the parallel plates\,\citep{Dzya,ParNin1969}. Related nanotechnology is emerging, exploiting both attractive and repulsive Casimir forces in different geometries and tailoring optical properties of different material combinations\,\citep{Bost2000,NetoPhysRevA,Zwol1,WoodsRevModPhys.88.045003,Ser2018,Woods2020,twosphere_Schoger_IntJModPhysA2022,MundayTorque2022,shelden2023enhanced,MundayReview2024,Woods2024}. In parallel,\,\cite{ZhangNature2024} found that magnetic fields can be used to manipulate the magnitude of the measured Casimir force. Notably, electrons and positrons can be generated from vacuum by magnetic fields above the Schwinger limit\,\citep{BREVIK2025139272}. 
The core of the present work is expanding the work by \,\cite{NinhamDaicicPhysRevA.57.1870} to account for the impact of a magnetic electron-positron plasma on Casimir forces at the nuclear scale. One question we addressed in the past was if nuclear Yukawa potentials\,\citep{Yukawa1935,Wick,Yukawa1955} receive a substantial contribution from Casimir forces\,\citep{PhysRevA.67.030701,EPJDNinham2014,Ninham_Brevik_Bostrom_2022}. In these studies, equivalence between vacuum fluctuations of quantized field and Yukawa interactions has been reported by studying contributions of the screened Casimir energy between nuclear particles in electron-positron sea to nuclear interactions and by interpreting mesons as plasmons in electron-positron plasma. Here, we derive general theory, present numerical results, and discuss potential implications. Our focus is on the impact of Casimir effects at the nuclear scale, and in particular, the effect of dielectric and magnetic properties of an electron-positron plasma on Casimir forces. We further explore the influence of a magnetic field on the screened Casimir force between two perfectly conducting surfaces across a magnetic electron-positron plasma. 
 
 \section{Theoretical Background}
 We discuss how the low-temperature
Casimir interaction  across vacuum
can generate, locally between very close surfaces, the effective temperature required to
give rise to an electron-positron plasma. 
When expanding the Casimir free energy between two perfect conducting surfaces (without plasma in between), the dominating terms in a series expansion correspond to (a) the zero-point Casimir energy at zero temperature; (b) a chemical potential contribution (identical to the energy of an high density electron-positron plasma\,\citep{Landau}); (c) the black body radiation in the gap. It has been suggested that Casimir effect can generate the high potentials needed to cause the high temperatures required for electron-positron generation\,\citep{Ninham_Brevik_Bostrom_2022}. The proposed source of high temperatures in the volume between two nuclear particles originates from assuming that the zero-point energy generates the black-body radiation energy.  
In the current work, we follow a slightly different, and more general, approach that accounts for magnetic permeability and explores the impact of magnetic fields. 

\subsection{Casimir Effect and Black-Body Energy Generated Plasma}
\label{CasBBEnergyGenPlasma}
  
We consider the physical system of two perfectly planar plates with an area corresponding to two protons. The energy density of blackbody radiation (i.e., electromagnetic radiation in thermal equilibrium at temperature $T$) is given by\,\citep{Landau}
\begin{equation}
u(T) = \frac{\pi^2}{15} \, \frac{(k_{\rm B} T)^4}{\hbar^3 c^3}. \label{Cas-Nucl-2}
\end{equation}
If the radiation fills a volume $V = A L$, where $A=\pi R^2$ is the area of the plates, $R$ is the proton radius, the total blackbody energy is
\begin{equation}
E_{\rm BB} = \left( \frac{\pi^2}{15} \, \frac{(k_{\rm B} T)^4}{\hbar^3 c^3} \right) \, V, \label{Cas-Nucl-3}
\end{equation}
where $k_{\rm B}$ is Boltzmann constant, and $T$ is absolute temperature, and $V$ is the volume given above of the blackbody. Now equaling the magnitudes of $E_C$ and $E_{\rm BB}$
\begin{equation}
\frac{\pi^2 \hbar c A}{720 L^3} = \frac{\pi^2 (k_B T)^4}{15 \hbar^3 c^3}  A L \label{Cas-Nucl-4}
\end{equation}
and we find
\begin{equation}
T = \frac{\hbar c}{ k_B }. \frac{1}{\gamma L}, \quad \text{where } \, \gamma = 48^{1/4}. \label{Cas-Nucl-5}
\end{equation}
Note that in the above equation temperature of black-body generated plasma explicitly dependent on the plate separation distance $L$. Using the expression of the Casimir force between the plates given in Eq.\,(\ref{Cas-Nucl-1}) we re-express the above equation as
\begin{equation}
T = \left( \frac{5 \hbar^3 c^3}{\pi^2 k_B^4 } \, \frac{F_C}{A} \right)^{1/4}. \label{Cas-Nucl-6}
\end{equation}

Next we start with the generic expression for the number density of fermions in thermal equilibrium
\begin{equation}
\rho = g \int \frac{d^3p}{(2\pi \hbar)^3} \, \frac{1}{e^{\beta E_p} + 1},
\end{equation}
where $g=2$ is the internal degrees of freedom for electron or positron, $\beta=1/k_{\rm B}T$,  and $E_{p}=\sqrt{p^2c^2+m^2c^4}$. At high temperature, considering $k_B T \gg m c^2$, we achieve $E_p = p c$, then we obtain
\begin{equation}
\rho = \frac{g}{(2\pi \hbar)^3} \, \int_0^\infty \frac{ 4\pi p^2 dp}{e^{\beta p c} + 1}.
\end{equation}
We rewrite the latter with the dimensionless variable $x=\beta p c$
\begin{equation}
\rho = \frac{4\pi g}{(2\pi \hbar)^3} \, \frac{1}{(\beta c)^3} \, \int_0^\infty \frac{x^2 dx}{e^x + 1}. \label{density-0}
\end{equation}
Identifying the Riemann-Zeta function,  $\displaystyle\int_0^\infty \frac{x^2 dx}{e^x + 1} = \frac{3}{2} \zeta(3)$, we find the number density to be (\,\cite{Landau})
\begin{equation}
\rho = \frac{3 g \zeta(3)}{4 \pi^2} \cdot \frac{(k_B T)^3}{\hbar^3 c^3}.
\end{equation}
Total number density for electrons and positrons is found to be
\begin{equation}
\rho=\rho_{e^-}+\rho_{e^+} = \frac{3 \zeta(3)}{\pi^2} \, \frac{(k_B T)^3}{\hbar^3 c^3} \label{no-density}
\end{equation}
Substituting Eq.\,(\ref{Cas-Nucl-5}) into the above
\begin{equation}
\rho= \frac{3 \zeta(3)}{\pi^2} \, \frac{1}{\hbar^3 c^3} \cdot \left( \frac{\hbar c}{\gamma L} \right)^3 \approx 0.020 \, \frac{1}{L^3}.
\label{Eq:total-number-density}
\end{equation}
Note here that total number density for electrons and positrons has explicit dependence on $L$, the distance between the two perfectly conducting surfaces.

\subsection{Magntitude of Casimir Force at Nuclear Scale}

How large is the Casimir force at a nuclear scale. This was first discussed in a not-yet-published manuscript by Ninham and Pask from 1969. Some estimates were given by \,\cite{RFleming} using the proximity force theorem for the Casimir force
for electrons and protons. This work indicated that Casimir and electrostatic interactions are similar in magnitude at the relevant length scales. For completeness, we proceed with the equilibrium between Casimir force and the Coloumb force. We consider two protons separated by a distance $L$, and each proton has an effective radius $R$. 
When the center of mass of two protons are separated by a distance of $2R + L$, the Coulomb interaction energy is approximated by\,\citep{landau1960electrodynamics}

\begin{equation}
E_{\text{Coulomb}} = \frac{1}{4\pi \epsilon_0}\, \frac{e^2}{2R + L}.
\end{equation}
where $e$ is the electron charge unit, and $\epsilon_0$ is the vacuum permittivity. We now equate the magnitudes of Casimir energy and Coloumb energy
\begin{equation}
\frac{\pi^3 \hbar c R^2}{720 L^3} = \frac{1}{4\pi \epsilon_0} \, \frac{e^2}{2R + L}
\end{equation}
Defining the dimensionless quantities  $\tilde x = \frac{L}{R}$ and $D = \frac{\pi^4 \epsilon_0 \hbar c }{180 e^2}$, we find
\begin{equation}
\tilde  x^3 - D \tilde  x - 2D = 0 .
\end{equation}
We solve the above equation using Cardano's method and find
\begin{equation}
\tilde  x = \sqrt[3]{ D + \sqrt{ D^2 - \frac{D^3}{27} } } + \sqrt[3]{ D - \sqrt{ D^2 - \frac{D^3}{27} } } \label{equilibrium}
\end{equation}
This gives us an equilibrium distance of 2.6\,fm. This result suggest that the Casimir force could be relevant for nuclear scale distances. 

\section{Magnetic Electron-Positron Plasma}

\subsection{Casimir Interaction in Electron-Positron Plasmas }

Consider the Casimir-Lifshitz interaction between  ideal metal surfaces separated by a plasma with dielectric permittivity
\begin{equation}
\varepsilon (\omega ) = 1 - \frac{{\rho {e^2}}}{{\epsilon_0 m{\omega ^2}}},
\label{Eq1}
\end{equation}
where  $\rho=\rho_{e^{-}}+\rho_{e^{+}}$ is the total number density of the electron-positron plasma given by Eq.(\ref{no-density}), and $m$ is the electron mass. We identify the plasma frequency $\omega_{ep}^2= \rho e^2/\epsilon_0 m$.

We have found that the asymptotic interaction free energy (for metal plates with area $A$) can at high temperatures and/or large separations
be written as\,\citep{PhysRevA.67.030701,EPJDNinham2014,Ninham_Brevik_Bostrom_2022},
\begin{align}
\mathcal{F}(L,T)&=\mathcal{F}_{n=0}(L,T)+\mathcal{F}_{n>0}(L,T),\label{Eq55}
\end{align}
where we separated the zero {$\mathcal{F}_{n=0}(L,T)$} and finite frequency {$\mathcal{F}_{n>0}(L,T)$} contributions, 
\begin{align}
\frac{\mathcal{F}_{n=0}(L,T)}{A} &\approx  - \frac{k_{\rm B}T}{2\pi}\,  \kappa^2\, e^{-2 {\kappa}\,  L} \left[\frac{1}{2 \kappa L} + \frac{1}{(2 \kappa L )^2}\right],
\label{Eq56}\\
\frac{{\mathcal{F}_{n > 0}}(L,T)}{A} &\approx -\frac{{{{(k_{\rm B}T)}^2}}}{\hbar c}\frac{e^{ - \pi \bar \rho \bar x}e^{ - 2\pi \bar x}}{L} + O({e^{ - {x^2}}}),
\label{Eq57}
\end{align}
with the convenient variables  $\kappa=\sqrt{\mu_{ep}}\,\omega_{ep}/c$,  $\bar x=2 k_{\rm B} T L/(\hbar c)$, and  $\bar \rho  = \rho {e^2}{\hbar ^2}/\left(4 \pi^2 m\epsilon_0 (k_{B}T)^2 \right)$. In these expressions, $\mu_{ep}$ is the magnetic permeability of electron-positron plasma, $k_{B}$ is Boltzmann's constant, $\hbar$ Planck's constant, $T$ effective temperature of the plasma, $c$ the velocity of light, and $L$ the distance between the plates. The dependence of the zero-frequency term on magnetic permeability will be explained in the following sections.

\subsection{Zero-Frequency Contribution to the Lifshitz Formula with Static Magnetic Permeability}

For two parallel planar media separated by another medium, the Lifshitz free energy per unit area is (\cite{lifshitz})
\be
\frac{\mathcal{F}(L) }{A}= k_{B} T \, \sum_{n=0}^{\infty}{}^\prime  \int \frac{d^2 \mathbf{k}_\perp}{(2\pi)^2} \sum_{\sigma} \log \left[1 - r_{12}^{\sigma}(i\xi_n)\, r_{32}^{\sigma}(i\xi_n) e^{-2 \kappa_2 L} \right], \label{Eq:lifshitz}
\ee
where the prime on the sum means the $n = 0$ term gets half weight, $\xi_n = {2\pi n k_{\rm B} T}/{\hbar}$ are the Matsubara frequencies. 
For planar structures the quantum number that characterizes the normal modes is $\sigma$, and it is related to the conservation of the $k_\parallel$. 
Two mode types can occur: transverse magnetic (TM) and transverse electric (TE), and they are related to the plane formed by the wave vector and each planar surface. In the above equation $L$ is the thickness of intermediate medium, and the  reflection
coefficients for a wave impinging on an interface between medium $i$ and $j$ from the $i$-side given as
\begin{equation}
    r_{ij}^{\mathrm{TM}} = \frac{{{\varepsilon _j}{\kappa _i} - {\varepsilon _i}{\kappa _j}}}{{{\varepsilon _j}{\kappa _i} + {\varepsilon _i}{\kappa _j}}},
    \label{eq:radialTM}
\end{equation}
and
\begin{equation}
    r_{ij}^{\mathrm{TE}} = \frac{{ {\mu_{j}{\kappa _i} - \mu_{i}{\kappa _j}} }}{{{\mu_{j}{\kappa _i} + \mu_{i}{\kappa _j}} }},
    \label{eq:radialTE}
\end{equation}
for TM and TE modes, respectively.
Here, $\kappa$ stands for the (imaginary) perpendicular component of the wave vector 
\begin{equation}
    {\kappa _i} (i \xi_n) = \sqrt{k_{\perp}^2 +{\varepsilon _i}\left(i\xi_{n}\right)\,{\mu_{i}}\left( i\xi_{n}  \right){{\left( {\xi_{n} /c} \right)}^2}},\label{eq:gamma}
\end{equation}
where ${{\varepsilon _i}\left(\omega\right)}$
is the dielectric function and ${\mu_{i}}\left( \omega  \right)$ is the magnetic permeability of medium $i$,  and $c$ the speed of light in
 vacuum. We assume two plasma surfaces with infinite plasma frequency interacting across an electron-positron plasma described in this work. The Lifshitz formula for the zero-frequency contribution reads

\begin{align}
    \frac{\mathcal{F}_{n=0}(L)}{A}
= \frac{{  k_{\rm B}T}}{{4\pi}}  &\int_0^\infty dk_{\perp}   k_{\perp} \times  \nonumber \\
&\left[ \log \Big(1-A^{\bf{TM}} \, e^{-2  \kappa_2 L }\Big)  +  \log \Big(1-A^{\bf{TE}} \, e^{-2 \kappa_2 L }\Big)\right],
\label{EqA9}
\end{align}
where  $A^{\bf{TM}}=r^{TM}_{12}r^{TM}_{32}\mid_{\xi=0}$  and $A^{\bf{TE}}=r^{TE}_{12}r^{TE}_{32}\mid_{\xi=0}$.  Thus we find
\bea
A^{\bf{TM}}&=&r_{12}^{TM}r_{32}^{TM}= \bigg(\frac{\varepsilon_{1}\kappa_{2}-\varepsilon_{2}\kappa_{1}} {\varepsilon_{1}\kappa_{2}+\varepsilon_{2}\kappa_{1}}\bigg)\bigg(\frac{\varepsilon_{3}\kappa_{2}-\varepsilon_{2}\kappa_{3}}{\varepsilon_{3}\kappa_{2}+\varepsilon_{2}\kappa_{3}}\bigg)\bigg{|}_{\xi=0} 
\eea
As $\varepsilon_1=\varepsilon_3$, from the above after calculational simplifications we obtain
\be
A^{\bf{TM}}= \left(\frac{\omega_{pc}^{2}\,\sqrt{c^2k_{\perp}^2 +\mu_{ep}\,\omega_{ep}^{2}}-\omega_{ep}^{2}\,\sqrt{c^2k_{\perp}^2 +\omega_{pc}^{2}}}{\omega_{pc}^{2}\,\sqrt{c^2k_{\perp}^2 +\mu_{ep}\,\omega_{ep}^{2}}+\omega_{ep}^{2}\,\sqrt{c^2k_{\perp}^2 +\omega_{pc}^{2}}}\right)^2 .
\ee
For the perfect conductor, or perfect metal plate, (we take the perfect conductor limit $\omega_{pc}$ to $\infty$ as $\varepsilon_{1}$ and $\varepsilon_{3}$ go to infinity for perfect metal plates. Thus we find
\begin{align}
A^{\bf{TM}} &= \lim_{\omega_{pc} \rightarrow \infty} \left(\frac{\sqrt{c^2k_{\perp}^2 +\mu_{ep}\,\omega_{ep}^{2}}-\frac{\omega_{ep}^{2}}{\omega_{pc}^2}\,\sqrt{\frac{c^2k_{\perp}^2}{\omega_{pc}^{2}} +1}}{\sqrt{c^2k_{\perp}^2 +\mu_{ep}\,\omega_{ep}^{2}}+\frac{\omega_{ep}^{2}}{\omega_{pc}^2}\,\sqrt{\frac{c^2k_{\perp}^2}{\omega_{pc}^{2}} +1}}\right)^2 
= 1.
\end{align}
Similarly, 
\begin{align}
    A^{\bf{TE}}&=r_{12}^{TE}r_{32}^{TE}= \bigg(\frac{\mu_{1}\kappa_{2}-\mu_{2}\kappa_{1}} {\mu_{1}\kappa_{2}+\mu_{2}\kappa_{1}}\bigg)\bigg(\frac{\mu_{3}\kappa_{2}-\mu_{2}\kappa_{3}}{\mu_{3}\kappa_{2}+\mu_{2}\kappa_{3}}\bigg)\bigg{|}_{\xi=0} \nonumber \\
    &=  \left[ \frac{   \sqrt{k_{\perp}^2 +\mu_2 \frac{\omega_{ep}^2}{c^2}}- \mu_2  \sqrt{k_{\perp}^2 + \frac{\omega_{pc}^2}{c^2}} }{   \sqrt{k_{\perp}^2 +\mu_2 \frac{\omega_{ep}^2}{c^2}}+ \mu_2  \sqrt{k_{\perp}^2 + \frac{\omega_{pc}^2}{c^2}} }\right]^2.
\end{align}
Now we take again the perfect conductor limit ($\omega_{pc} \rightarrow \infty$) and we find $ A^{\bf{TE}} ~=~ 1$. 
Thus from Eq.\,(\ref{EqA9}), we find
\begin{equation}
\frac{\mathcal{F}_{n=0}(L)}{A} = \frac{{  k_{\rm B}T}}{{2\pi}}  \int_0^\infty dk_{\perp}   k_{\perp}  \log \Big(1-e^{-2  \kappa_2 L }\Big),
\label{EqA9a}
\end{equation}
where $\kappa_{2}=\sqrt{k_{\perp}^2 +\mu_{ep}\,\frac{\omega_{ep}^{2}}{c^2}}=\sqrt{k_{\perp}^2 +{\kappa}^2}$,  $\mu_{ep}$ is the static magnetic permeability of the electron-positron plasma, and $\omega_{ep}^2= \rho e^2/\epsilon_0 m$.
Note that the above equation lead to Eq.\,(\ref{Eq56}) at high temperature and/or large separation. The magnetic permeability will be described in the following section.

\subsection{Magnetic Permeability of an Electron-Positron Plasma}

Consider a paramagnetic material containing $N$ magnetic dipoles per unit volume with a permanent magnetic moment $\bar{\mu}$ per each dipole. When we expose these dipoles to an external magnetic field $\vec{B}$, it starts precision over themselves in the direction of the field. Thermal energy at room temperature opposes this orientation and in thermal equilibrium, dipoles orient with angle $\theta$ with the direction of magnetic field. Now the interaction energy of a magnetic dipole in the field is (\cite{Kunz})
\be
E =  -\bar{\mu} B \cos\theta.
\ee
Now using the Boltzmann distribution law, we find that the number of magnetic dipoles having this orientation (probability distribution of this orientation) is 
\be
n(\theta) \propto e^{\bar{\mu} B \cos\theta/k_{\rm B} T}
\ee
The probability for a magnetic dipole to make an angle between $\theta$ and $\theta +d\theta$ with the magnetic field  is given by
\be 
dn(\theta) \propto  e^{\bar{\mu} B \cos\theta/k_{\rm B} T} d\omega .
\ee
The average component of the magnetic moment of each atom along the field direction multiplied by the number of dipoles per unit volume, N, gives the magnetization
\be
M=N\langle \bar{\mu} \cos\theta \rangle = N\frac{\int_{0}^{\pi} \bar{\mu} \cos\theta \,\sin\theta e^{\frac{\bar{\mu} B \cos\theta}{k_{\rm B} T}} \, d\theta}{\int_{0}^{\pi} \sin\theta e^{\frac{\bar{\mu} B \cos\theta}{k_{\rm B} T}} \, d\theta} .
\ee
Using redefinition $x = \cos\theta$, we find
\be
\langle \bar{\mu} \cos\theta \rangle = \bar{\mu} \frac{\int_{1}^{-1} x e^{y x} dx}{\int_{1}^{-1} e^{y x} dx}, \quad y = \frac{\bar{\mu} B}{k_{\rm B} T} \label{Langevin-0} .
\ee
Using $\int_{1}^{-1} e^{\alpha x} dx = \frac{2\sinh\alpha}{\alpha}$ and $\int_{1}^{-1} x e^{\alpha x} dx = \frac{2(\alpha \cosh\alpha - \sinh\alpha)}{\alpha^2}$ in the above equation we find
\be
M=N\langle \bar{\mu} \cos\theta \rangle = N \bar{\mu} \left( \coth y - \frac{1}{y} \right) = N \bar{\mu} \mathcal{L}(y),
\label{Langevin}
\ee
where $\mathcal{L}(y)$ is the Langevin function. Now for $y \ll 1$ (small magnetic fields and/or high temperatures), $\mathcal{L}(y) \approx y/3$. Thus we find
\be
M \approx N \bar{\mu} \left( \frac{y}{3} \right) = \frac{N\bar{\mu}^2 B}{3k_{\rm B} T}  
\ee
From above equation using $B \approx \mu_0 H$, we find magnetic susceptibility and magnetic permeability (relative) as
\be
\chi = \frac{\partial M}{\partial H} = \mu_0 \frac{\partial M}{\partial B} = \frac{\mu_0 N \bar{\mu}^2}{3k_{\rm B} T} ~~and~~ \mu_{r}=1+\frac{\mu_0 N \bar{\mu}^2}{3k_{\rm B} T}.
\label{susceptibility_permeability}
\ee
Note that here $\bar{\mu}$ intrinsic magnetic moment. Now  we see the quantum mechanical prescription of finding magnetic susceptibility and relative permeability. As pointed out by\,\cite{ZhangNature2024}
 Eq.\,(\ref{Langevin}) and Eq.\,(\ref{susceptibility_permeability}) give zero magnetic susceptibility in the limit of high magnetic fields.  In the numerical section, we explore the impact of magnetic fields on the magnitude of nuclear scale Casimir potentials, screening length, and the quantity corresponding to the effective Casimir "meson mass".
We start with the magnetic moment associated with each atom in a paramagnet is given by (\cite{van1932theory, Kittel})
\be
\bar{\mu} = -g \mu_{\rm B} \sqrt{J(J+1)},
\ee
where $\mu_{\rm B}=\frac{e\hbar}{2m}$  is the Bohr magneton and $g$ is the Landé g-factor defined as 
\be
g = 1 + \frac{J(J + 1) + S(S + 1) - L(L + 1)}{2J(J + 1)},
\ee
where $S$ is spin angular momentum quantum number, $L$ orbital angular momentum quantum number and $J$ is total angular momentum quantum number.

In the presence of an external magnetic field $B$, the energy levels split according to the magnetic moment orientation. The orientation of magnetic moment $\bar{\mu}$ with respect to the direction of the applied magnetic field are specified components of $\bar{\mu}$ along the field direction is
\be 
\bar{\mu}_{z}=-g\mu_{\rm B}m_{\rm J},
\ee
where $m_{\rm J}=-J, -J+1,......J-1, J$ is the magnetic quantum number associated with total angular momentum $\textbf{J}$. For each value of J, $m_{J}$ can have $(2J+1)$ values. Magnetic moment has possible $(2J+1)$ different orientation relative to the field. Now the interaction energy of a magnetic dipole in presence of  $B$, given  by
\be 
E=m_{\rm J}g \mu_{\rm B}B.
\ee
According to the Boltzmann distribution, the number of magnetic dipole having particular $m_{\rm J}$ is propotional to 
\be 
\exp\left(-\frac{g\mu_{\rm B} B m_{\rm J}}{K_{\rm B}T}\right),
\ee
and then, the magnetization
\be
M = N \langle \bar{\mu} \rangle =  N \sum_{m_{\rm J}=-J}^{m_{\rm J}=+J}\frac{ g \mu_{\rm B} m_{\rm J} e^{-\beta E}}{e^{-\beta E}} .
\ee
After performing the summation over $ m_{\rm J} $ and after some calculational simplification we obtain the magnetic susceptibility as
\be
\chi = \mu_{0}\frac{M}{B}=\frac{N \mu_0\, g^2 \mu_{\rm B}^2 J(J+1)}{3 k_{\rm B} T}
\ee
and magnetic permeability $\mu_{r}=1+\chi$, i.e.,
\be 
\mu_{r}=1+\frac{N \mu_0\, g^2 \mu_{\rm B}^2 J(J+1)}{3 k_{\rm B} T}.
\ee
Now to find the magnetic susceptibility and permeability for electron we take $ S= \frac{1}{2}$,  $L = 0$ and thus $J = \frac{1}{2}$. Lande' g-factor $g=2$. Using these details we find
\be
\chi_{e^{-}} = \frac{\mu_0 N g^2 \mu_{\rm B}^2 J(J+1)}{3 k_{\rm B} T} = \frac{\mu_0 N \mu_{\rm B}^2}{k_{\rm B} T}
\ee
and
\be
\mu_{e^{-}} = 1 + \chi_{e^{-}} = 1 + \frac{\mu_0 N \mu_{\rm B}^2}{k_{\rm B} T} .
\ee
As the positron has the same spin as the electron and opposite charge of electron (i.e $\mu_{\rm B}^2$ will give same value for electron and positron), thus we find magnetic susceptibility
\be
\chi_{e^{+}} = \chi_{e^{-}} = \frac{\mu_0 N \mu_{\rm B}^2}{k_{\rm B} T}.
\ee
Similarly we find the permeability for positron as
\be
\mu_{e^{+}} = 1 + \chi_{e^{+}} = 1 + \frac{\mu_0 N \mu_{\rm B}^2}{k_{\rm B} T}.
\ee
In plasma we consider $N_{e^{-}} = N_{e^{+}} = N$\,\citep{10.1063/5.0156153} and considering both species will contribute additively in the calculation of magnetic susceptibility for an electron-positron plasma. Thus we find
\be
\chi_{(e^{-},e^{+})} = \chi_{e^{-}} + \chi_{e^{+}} = \frac{2 \mu_0 N \mu_{\rm B}^2}{k_{\rm B} T}
\ee
and permeability at zero frequency
\be
\mu_{ep} = 1 + \chi_{(e^{-},e^{+})} = 1 + \frac{2 \mu_0 N \mu_{\rm B}^2}{k_{\rm B} T}. \label{permeability}
\ee
Note that in the above equation we identify $\rho={\rho _{e^{-}} } + {\rho _{e^{+}}}=2N$. The generalization for finite frequencies behave as
\begin{equation}
\mu_{ep}(i \xi_n)=1+\frac{2 \mu_0 N \mu_{\rm B}^2}{k_{\rm B}T (1+\xi_n^2/\omega_{\mu}^2)}.\label{Eq:mu-ep}
\end{equation}
Here $\omega_{\mu}$ is relatively low, typically a magnetic proper frequency is of the order $10^{10}$ rad/s.
Since for high temperatures $\mu_{ep}(0)-1<<(1+\xi_1^2/\omega_{\mu}^2)$ finite frequency permeability is very close to 1.  In the presence of a magnetic field, the zero-frequency magnetic permeability can, in the case of an electron-positron plasma, be estimated as 
\be
\mu_{ep}(H) = 1 + \frac{6N \bar{\mu}}{H}\,\mathcal{L}(y) ,
\ee
where $y$ is given in Eq.\,(\ref{Langevin-0}) and $\mathcal{L}(y)$ is the Langevin function given in Eq.\,(\ref{Langevin}). In the above equation if $y \ll 1$ it represents small magnetic fields and/or high temperatures and this lead to $\mathcal{L}(y) \approx y/3$. Using this we get back the Eq.\,(\ref{permeability}). However, in the opposite limit, magnetic effects are important with the magnetic permeability approaching unity for high magnetic field strengths. This could, potentially, lead to some interesting predictions in astroparticle physics such as modifications in properties and dynamics of the magnetized electron–positron plasma in the relativistic jets around Pulsars and Neutron stars\,\citep{Homan_2009,KUMAR20151,  Stoneking_Pedersen_Helander}, in the early Universe\,\citep{GRAYSON2023169453}. 
\begin{figure}[!h]
    \centering
    \includegraphics[width=1\linewidth]{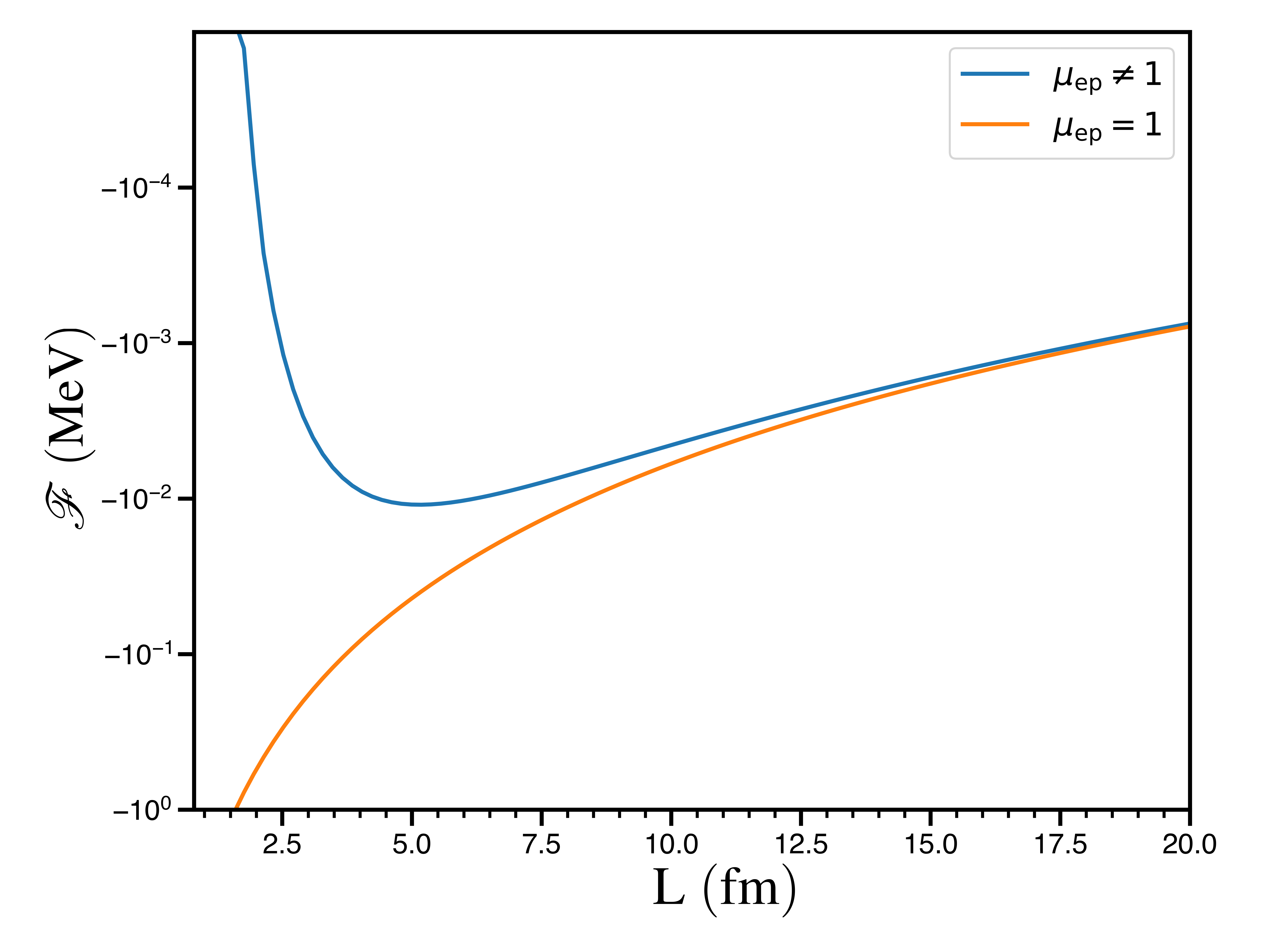}
    \vspace{-0.2in}
    \caption{Zero frequency contribution to interaction free energy from Eq.(\ref{Eq56}), considering either $\mu_{ep}=1$ (lower orange curve) or Eq.(\ref{permeability}) (upper blue curve). The x-axis is the $L$ distance in femtometers, y-axis is the Casimir interaction free energy in MeV for two perfect conducting plates with an area ($A=\pi R^2$) given in the text and an intermediate plasma density varying with distance via the equilibrium of zero temperature Casimir energy and the black body radiation energy.  }
    \label{fig:graph1}
\end{figure}

\section{Numerical Results: Impact from Magnetic Permeability}
\label{NumResSec}

Many of the results presented above are general and valid for Casimir effect across any high-density magnetic plasma. In this section, we provide numerical results and, in addition, speculate about physical impacts at the nuclear scale. 
We compare in Fig.\,\ref{fig:graph1} the zero frequency contribution when the permeability is one or given by Eq.(\ref{permeability}). There are important magnetic corrections to the Casimir interaction free energy. Remarkably, the interactions are of the same order of magnitude as both the nuclear and the Coulomb interactions. 
Notably, since the first finite Matsubara frequency is high the magnetic effects only impact the zero-frequency term. Our purpose in this work has been to show how simple it is to include the previously neglected magnetic contribution to the Casimir force between nuclear surfaces across an electron-positron plasma. We summarize some of the main results in the equations below and in Table\,\ref{Table:temperature-distance}.
\begin{align}
    \kappa &= \sqrt{\mu_{ep}} \frac{\omega_{ep}}{c}= \frac{\sqrt[8]{3}}{2\pi} \sqrt{\frac{e^2\mu_0\zeta(3)}{2L^3 m} \left( 1+\frac{\sqrt{3}e^2\hbar \mu_0 \zeta(3)}{8 \pi^2 c L^2 m^2 }\right)}\\
    \frac{\mathcal{F}_{n=0}}{A}&=  - \frac{\hbar c }{4 \sqrt[4]{3}\, \pi L}\,  \kappa^2\, e^{-2 {\kappa}\,  L} \left[\frac{1}{2 \kappa L} + \frac{1}{(2 \kappa L )^2}\right],\\
    \frac{\mathcal{F}_{n>0}}{A}&=-\frac{\hbar c}{4 \sqrt{3} L^3}\exp \left(\frac{-\sqrt{3}\, \zeta(3) e^2 \mu_0}{8 \pi^ 3 m L }-\frac{2\pi}{\sqrt[4]{3}}\right).
\end{align}
The last expression, for the finite frequency contribution, is only used when considering the case where temperature is allowed to vary with distance. In all other cases, we use Eq.(\ref{Eq57}) and fix the temperature starting from an initial plate separation where the temperature is assumed to be generated by the zero-temperature Casimir force between two perfectly reflecting plates across a quantum vacuum.

\begin{table}[h]
\centering
\begin{tabular}{|c|c|c|}
\hline
\textbf{Quantity/Variable}     & \textbf{Temperature}                                                & \textbf{Distance}                                           \\ \hline \hline
Temperature                    &                                                                 T    & $\frac{\hbar c}{\sqrt[4]{48}\, k_{\rm B} L}$ \\ \hline
Density    $\rho$                    & $\frac{3 \zeta(3)}{\pi^2} \left(\frac{k_{\rm B}T}{\hbar c}\right)^3$ & $\frac{\zeta(3)}{8 \pi^2 L^ 3}\sqrt[4]{3}$                                    \\ \hline
Plasma freq. $\omega_{ep}$ & $\sqrt{\frac{\rho(T) e^2}{\epsilon_0 m}}$                           & $\frac{\sqrt[8]{3}}{2\pi} \sqrt{\frac{e^2 \zeta(3)}{2 m \epsilon_0 L^3}}$                               \\ \hline Permeability $\mu_{ep}$        & $1+\frac{2\mu_0 \rho (T)\mu_B^2}{k_{\rm B}T}$                               & $1+\frac{\sqrt{3} \mu_0 e^2 \hbar \zeta(3)}{8 \pi^2  L^2m^2c}$ \\ \hline
\end{tabular}
\caption{Functions in terms of temperature and distance.}
\label{Table:temperature-distance}
\end{table}

\begin{table}[h]
\centering
\begin{tabular}{|c|c|c|c|c|}
\hline
Distance & T   (K)                  & $\rho$ ($m^{-3}$)              & $\omega_{ep}$ (rad/s)       & $\mu_{ep}$ \\ \hline\hline
1\,fm      & $8.70\times 10^{11} $ & $2\times 10^{43}$   & $2.5\times 10^{23}$ & 360.8      \\ \hline
1.5\,fm    & $5.80\times 10^{11} $ & $5.9\times 10^{42}$ & $1.4\times 10^{23}$ & 160.9      \\ \hline
2\,fm      & $4.35\times 10^{11} $ & $2.5\times 10^{42}$ & $8.9\times 10^{22}$ & 91.0       \\ \hline
2.6\,fm      & $3.36\times 10^{11} $ & $11.40\times 10^{41}$ & $6.02\times 10^{22}$ & 54.2   \\ \hline
3\,fm      & $2.90\times 10^{11} $ & $7.4\times 10^{41}$ & $4.9\times 10^{22}$ & 41.0       \\ \hline
\end{tabular}
\caption{Temperature, number density, plasma frequency, and magnetic permeability for different surface distances using the theoretical results given in Table\,\ref{Table:temperature-distance}.}
\label{Table:numeric-distance}
\end{table}

\begin{figure}[!ht]
    \centering
    \includegraphics[width=1\linewidth]{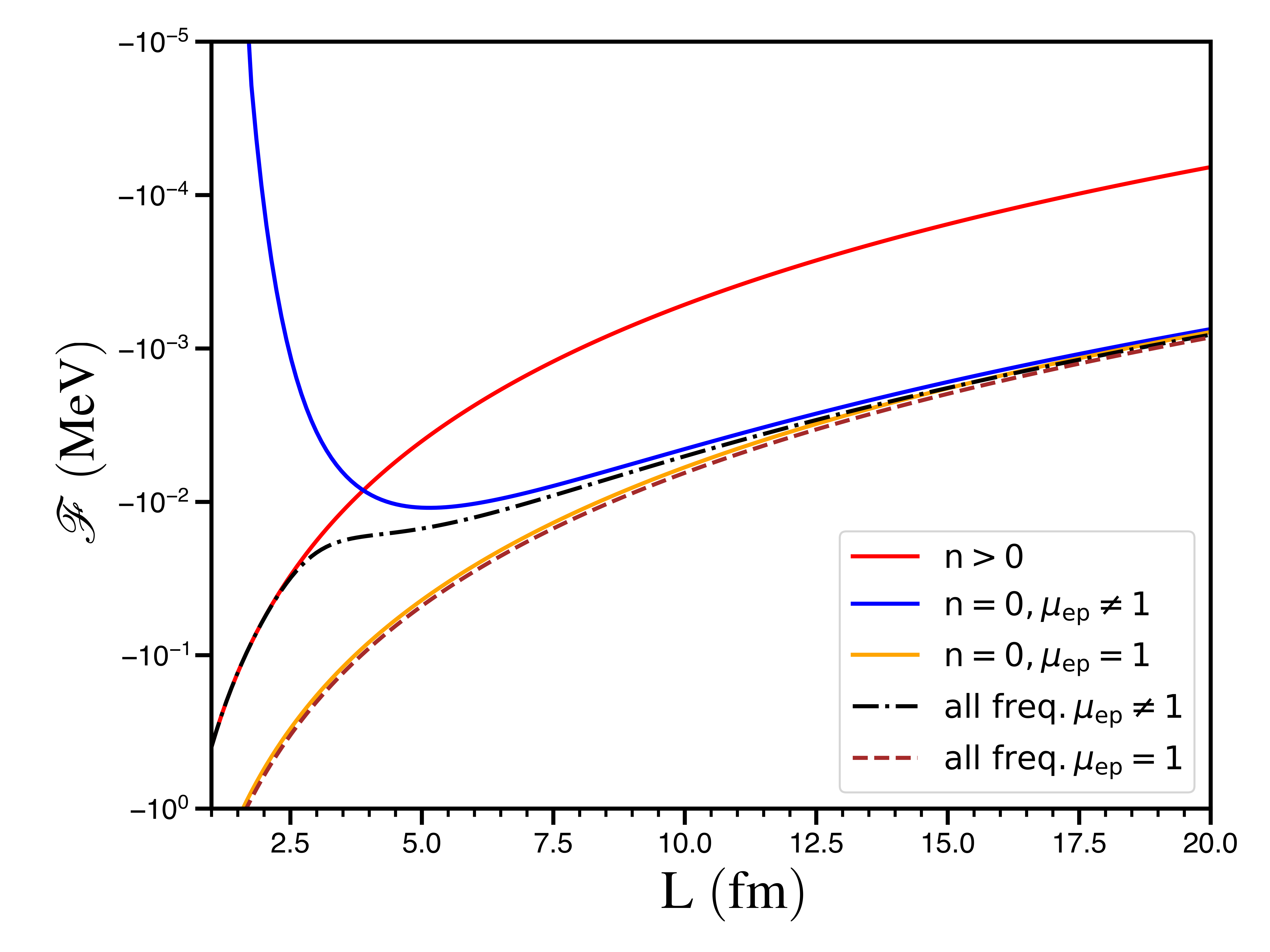}
     \vspace{-0.2in}
    \caption{Contributions from zero-frequency term, Eq.\,(\ref{Eq56}) (red),  and finite frequency terms, Eq.\,(\ref{Eq57}) (blue), and their sum (dashed black). The x-axis is the $L$ distance in femtometers, y-axis is the Casimir interaction free energy in MeV. In these examples, we consider two perfect conducting plates with an area ($A=\pi R^2$) given in the text and an intermediate plasma density varying with distance via the equilibrium of zero temperature Casimir energy and the black body radiation energy.  }
    \label{fig:graph2}
\end{figure}

In Fig.\,\ref{fig:graph2}, we explore different contributions to the interaction free energy for two perfect conducting plates with a surface area given in Sec.\,\ref{CasBBEnergyGenPlasma}. All contributions have temperatures varying with distance and include both the finite frequency term and the zero-frequency term. In the Table\,\ref{Table:numeric-distance} we present numerical values of temperature, number density, plasma frequency, and magnetic permeability at different separations, including the equilibrium surface distance $2.6$\,fm obtained from Eq.\,(\ref{equilibrium}).

\section{Discussions}

\,\cite{Feynman_PhysRev.80.440} reflected that {\it{"High energy potentials could excite states corresponding to other eigenvalues, possibly thereby corresponding to other masses. This note is meant only as a speculation, for not enough work has been done in this direction to make sure that a reasonable physical theory can be developed along these lines. (What little work has been done was not promising)"}}. We do not suggest that the Casimir-Yukawa theory can replace the well-explored theories of nuclear interactions. However, it is plausible that Casimir physics plays a small complementary role at the nuclear scale. With this in mind, we finally check how a quantity that behaves similarly to the meson mass depends on distance and spin magnetic permeability. The screened Casimir free energy asymptotes can be compared with the Yukawa potential between nuclear particles  at distances large compared with the screening length  $L_\pi=\hbar/m_\pi c$ (\cite{Yndurain}) where $m_\pi$  is the quantity that corresponds to the "mass of the mediating meson" in Yukawa theory,
\begin{equation}
\mathcal{F}(L,T) \propto e^{-L/L_\pi}.
\label{Eq58}
\end{equation}
The meson mass, if Casimir-Yukawa contribution for a second is assumed to be related to the Yukawa theory, is estimated  by taking the exponents in the $\mathcal{F}_{n=0}$ term  and the Yukawa potential to be equal
\begin{equation}
{m_\pi } = \frac{{2e\hbar }}{{{c^2}}}\sqrt {{\frac {\mu_{ep} \, ({\rho _{e^{-}} } + {\rho _{e^{+}}})} {\varepsilon_{0} m}}}.
\label{Eq59}
\end{equation}				
Since we know the meson mass (135 MeV), we can estimate the screening length in a magnetic and non-magnetic plasma, as well as in the presence of magnetic fields. The predicted ``masses'' for plates separated 1-3\,Fermi are 329 - 13\,MeV without magnetic effects included. The corresponding "masses" when we include magnetic (spin) permeability are 6242 - 84\,MeV.
These are just simple estimates, and we stress that mesons in general are not physically confined to only exist between two nuclear surfaces.

We have investigated screened intermolecular forces and would, eventually, like to touch on future extensions to the related lifetimes of any collective excitations of the electron-positron plasma. We can estimate the lifetimes of any electron-positron plasmons for the decay of a plasmon into two electron-positron pairs. These can subsequently decay, e.g., to produce two photons. The broadening ($\Delta E$) of an plasmon peak and its lifetime ($\tau\geq1/E$) is known analytically, and has been experimentally verified\,\citep{NinhamPhysRev.145.209},
\be
\Delta E \sim \frac{6\pi\varepsilon_{F}}{5\hbar} \Bigg(\frac{q_\pi}{q_F}\Bigg)^2\ \Bigg(\frac{\hbar\omega_p}{2\varepsilon_F}\Bigg)^3\ \Bigg[10\ln(2)+ 2 -4.5\frac{\hbar \omega_p}{2\varepsilon_F}+ 
\mathcal{O}\Bigg(\frac{\hbar \omega_p}{2\varepsilon_F}\Bigg)^2..\Bigg], 
\ee
The entities involved are the Fermi energy ($\varepsilon_F\propto\rho^{2/3}$), plasma frequency ($\omega_p\propto\rho^{1/2})$, and Fermi wavevector ($q_F\propto\rho^{1/3})$. These depend on density and hence on the distance between the two surfaces. The lifetime dependence upon the electron-positron plasma density can be deduced once we have a model for the neutral plasmon $q_\pi$-vector. Our planned expansions involve two-sphere geometry and, in addition, explore relativistic spin-dependent Casimir-Lifshitz interactions with plasmon excitations.

\section{Conclusions}

To conclude, we have explored how Casimir physics may play a non-negligible role at the nuclear scale. We have previously shown that a Casimir-Yukawa interaction potential at nuclear distances can be derived from a semi-classical theory in the presence of a non-magnetic electron-positron plasma\,\citep{PhysRevA.67.030701,EPJDNinham2014,Ninham_Brevik_Bostrom_2022}. In these past studies the magnetic effects were ignored.  In the current study, we examined the combined influence of dielectric and magnetic properties of an electron-positron plasma on Casimir forces.  
Our results are of the same order of magnitude as the experimentally found nuclear interaction energies that vary with the environment\,\citep{NuclearBindingAubertPLB1983,NuclearBinding_CLAS2019} (typically of the order 1.1\,MeV for deuterium to
8.8\,MeV for Nickel-62). As demonstrated by\,\cite{Ninham_Brevik_Bostrom_2022} the explicit form of the interaction with a plasma in the gap is the same as that for the Klein-Gordon--Yukawa potential. Our work shows how this form is modified when we account for the magnetic (spin) susceptibility.

\section*{Acknowledgements}
The research is part of the project No. 2022/47/P/ST3/01236
    co-funded by the National Science Centre and the European Union's Horizon 2020
    research and innovation programme under the Marie Sk{\l}odowska-Curie grant agreement
    No. 945339.  The research took place at the "ENSEMBLE3-Center of Excellence for nanophononics, advanced materials and novel crystal growth-based  technologies" project (GA No. MAB/2020/14) carried out under the International Research Agenda programs of the Foundation for Polish Science
    that are co-financed by the European Union under the European Regional  Development Fund and the European Union Horizon 2020 research and innovation program Teaming for Excellence (GA. No. 857543) for supporting this work. Our research contributions to this publication were created as part of the project
    of the Minister of Science and Higher Education "Support for the activities    of Centers of Excellence established in Poland under the Horizon 2020    program" under contract No. MEiN/2023/DIR/3797. SP acknowledge support from the Marie Sk{\l}odowska-Curie Doctoral Network TIMES, grant No. 101118915.

\section*{Data availability statement}

Upon special request, the data and code can be retrieved from the corresponding authors.

\section*{Conflict of interest }

The authors declare no Conflict of interest.

\bibliographystyle{elsarticle-harv} 

\begin{thebibliography}{40}
\expandafter\ifx\csname natexlab\endcsname\relax\def\natexlab#1{#1}\fi
\providecommand{\url}[1]{\texttt{#1}}
\providecommand{\href}[2]{#2}
\providecommand{\path}[1]{#1}
\providecommand{\DOIprefix}{doi:}
\providecommand{\ArXivprefix}{arXiv:}
\providecommand{\URLprefix}{URL: }
\providecommand{\Pubmedprefix}{pmid:}
\providecommand{\doi}[1]{\href{http://dx.doi.org/#1}{\path{#1}}}
\providecommand{\Pubmed}[1]{\href{pmid:#1}{\path{#1}}}
\providecommand{\bibinfo}[2]{#2}
\ifx\xfnm\relax \def\xfnm[#1]{\unskip,\space#1}\fi
\bibitem[{Aubert et~al.(1983)Aubert, Bassompierre, Becks, Best, Böhm, {de Bouard}, Brasse, Broll, Brown, Carr, Clifft, Cobb, Coignet, Combley, Court, D'Agostini, Dau, Davies, Déclais, Dobinson, Dosselli, Drees, Edwards, Edwards, Favier, Ferrero, Flauger, Gabathuler, Gamet, Gayler, Gerhardt, Gössling, Haas, Hamacher, Hayman, Henckes, Korbel, Landgraf, Leenen, Maire, Minssieux, Mohr, Montgomery, Moser, Mount, Norton, McNicholas, Osborne, Payre, Peroni, Pessard, Pietrzyk, Rith, Schneegans, Sloan, Stier, Stockhausen, Thénard, Thompson, Urban, Villers, Wahlen, Whalley, Williams, Williams, Williamson and Wimpenny}]{NuclearBindingAubertPLB1983}
\bibinfo{author}{Aubert, J.}, \bibinfo{author}{Bassompierre, G.}, \bibinfo{author}{Becks, K.}, \bibinfo{author}{Best, C.}, \bibinfo{author}{Böhm, E.}, \bibinfo{author}{{de Bouard}, X.}, \bibinfo{author}{Brasse, F.}, \bibinfo{author}{Broll, C.}, \bibinfo{author}{Brown, S.}, \bibinfo{author}{Carr, J.}, \bibinfo{author}{Clifft, R.}, \bibinfo{author}{Cobb, J.}, \bibinfo{author}{Coignet, G.}, \bibinfo{author}{Combley, F.}, \bibinfo{author}{Court, G.}, \bibinfo{author}{D'Agostini, G.}, \bibinfo{author}{Dau, W.}, \bibinfo{author}{Davies, J.}, \bibinfo{author}{Déclais, Y.}, \bibinfo{author}{Dobinson, R.}, \bibinfo{author}{Dosselli, U.}, \bibinfo{author}{Drees, J.}, \bibinfo{author}{Edwards, A.}, \bibinfo{author}{Edwards, M.}, \bibinfo{author}{Favier, J.}, \bibinfo{author}{Ferrero, M.}, \bibinfo{author}{Flauger, W.}, \bibinfo{author}{Gabathuler, E.}, \bibinfo{author}{Gamet, R.}, \bibinfo{author}{Gayler, J.}, \bibinfo{author}{Gerhardt, V.}, \bibinfo{author}{Gössling, C.}, \bibinfo{author}{Haas, J.},
  \bibinfo{author}{Hamacher, K.}, \bibinfo{author}{Hayman, P.}, \bibinfo{author}{Henckes, M.}, \bibinfo{author}{Korbel, V.}, \bibinfo{author}{Landgraf, U.}, \bibinfo{author}{Leenen, M.}, \bibinfo{author}{Maire, M.}, \bibinfo{author}{Minssieux, H.}, \bibinfo{author}{Mohr, W.}, \bibinfo{author}{Montgomery, H.}, \bibinfo{author}{Moser, K.}, \bibinfo{author}{Mount, R.}, \bibinfo{author}{Norton, P.}, \bibinfo{author}{McNicholas, J.}, \bibinfo{author}{Osborne, A.}, \bibinfo{author}{Payre, P.}, \bibinfo{author}{Peroni, C.}, \bibinfo{author}{Pessard, H.}, \bibinfo{author}{Pietrzyk, U.}, \bibinfo{author}{Rith, K.}, \bibinfo{author}{Schneegans, M.}, \bibinfo{author}{Sloan, T.}, \bibinfo{author}{Stier, H.}, \bibinfo{author}{Stockhausen, W.}, \bibinfo{author}{Thénard, J.}, \bibinfo{author}{Thompson, J.}, \bibinfo{author}{Urban, L.}, \bibinfo{author}{Villers, M.}, \bibinfo{author}{Wahlen, H.}, \bibinfo{author}{Whalley, M.}, \bibinfo{author}{Williams, D.}, \bibinfo{author}{Williams, W.}, \bibinfo{author}{Williamson, J.},
  \bibinfo{author}{Wimpenny, S.}, \bibinfo{year}{1983}.
\newblock \bibinfo{title}{The ratio of the nucleon structure functions f2n for iron and deuterium}.
\newblock \bibinfo{journal}{Phys. Lett. B} \bibinfo{volume}{123}, \bibinfo{pages}{275--278}.
\newblock \URLprefix \url{https://www.sciencedirect.com/science/article/pii/0370269383904379}, \DOIprefix\doi{https://doi.org/10.1016/0370-2693(83)90437-9}.
\bibitem[{Bostr\"om and Sernelius(2000)}]{Bost2000}
\bibinfo{author}{Bostr\"om, M.}, \bibinfo{author}{Sernelius, B.E.}, \bibinfo{year}{2000}.
\newblock \bibinfo{title}{{Thermal Effects on the Casimir Force in the 0.1-5\,$\mu$\,m Range}}.
\newblock \bibinfo{journal}{Phys. Rev. Lett.} \bibinfo{volume}{84}, \bibinfo{pages}{4757}.
\newblock \DOIprefix\doi{10.1103/PhysRevLett.84.4757}.
\bibitem[{Brevik et~al.(2025)Brevik, Chaichian and Tureanu}]{BREVIK2025139272}
\bibinfo{author}{Brevik, I.H.}, \bibinfo{author}{Chaichian, M.M.}, \bibinfo{author}{Tureanu, A.}, \bibinfo{year}{2025}.
\newblock \bibinfo{title}{Below the schwinger critical magnetic field value, quantum vacuum and gamma-ray bursts delay}.
\newblock \bibinfo{journal}{Physics Letters B} \bibinfo{volume}{861}, \bibinfo{pages}{139272}.
\newblock \URLprefix \url{https://www.sciencedirect.com/science/article/pii/S0370269325000322}, \DOIprefix\doi{https://doi.org/10.1016/j.physletb.2025.139272}.
\bibitem[{Casimir(1948)}]{CH-0-HBG}
\bibinfo{author}{Casimir, H.B.G.}, \bibinfo{year}{1948}.
\newblock \bibinfo{title}{On the attraction between two perfectly conducting plates}.
\newblock \bibinfo{journal}{Proc. K. Ned. Akad. Wet.} \bibinfo{volume}{51}, \bibinfo{pages}{793}.
\bibitem[{Dzyaloshinskii et~al.(1961)Dzyaloshinskii, Lifshitz and Pitaevskii}]{Dzya}
\bibinfo{author}{Dzyaloshinskii, I.}, \bibinfo{author}{Lifshitz, E.}, \bibinfo{author}{Pitaevskii, L.}, \bibinfo{year}{1961}.
\newblock \bibinfo{title}{{The general theory of van der Waals forces}}.
\newblock \bibinfo{journal}{Adv. Phys.} \bibinfo{volume}{10}, \bibinfo{pages}{165--209}.
\newblock \DOIprefix\doi{10.1080/00018736100101281}.
\bibitem[{Feynman(1950)}]{Feynman_PhysRev.80.440}
\bibinfo{author}{Feynman, R.P.}, \bibinfo{year}{1950}.
\newblock \bibinfo{title}{Mathematical formulation of the quantum theory of electromagnetic interaction}.
\newblock \bibinfo{journal}{Phys. Rev.} \bibinfo{volume}{80}, \bibinfo{pages}{440--457}.
\newblock \URLprefix \url{https://link.aps.org/doi/10.1103/PhysRev.80.440}, \DOIprefix\doi{10.1103/PhysRev.80.440}.
\bibitem[{Fleming(2015)}]{RFleming}
\bibinfo{author}{Fleming, R.}, \bibinfo{year}{2015}.
\newblock \bibinfo{title}{{The Zero-Point Universe}}.
\newblock \bibinfo{publisher}{Createspace Independent Publishing Platform}.
\bibitem[{Frolov(2023)}]{10.1063/5.0156153}
\bibinfo{author}{Frolov, A.M.}, \bibinfo{year}{2023}.
\newblock \bibinfo{title}{Thermodynamics of the electron–positron plasma at very high temperatures and sources of annihilation γ -quanta in our galaxy}.
\newblock \bibinfo{journal}{Physics of Plasmas} \bibinfo{volume}{30}, \bibinfo{pages}{102701}.
\newblock \URLprefix \url{https://doi.org/10.1063/5.0156153}, \DOIprefix\doi{10.1063/5.0156153}.
\bibitem[{Grayson et~al.(2023)Grayson, Yang, Formanek and Rafelski}]{GRAYSON2023169453}
\bibinfo{author}{Grayson, C.}, \bibinfo{author}{Yang, C.T.}, \bibinfo{author}{Formanek, M.}, \bibinfo{author}{Rafelski, J.}, \bibinfo{year}{2023}.
\newblock \bibinfo{title}{Electron–positron plasma in bbn: Damped-dynamic screening}.
\newblock \bibinfo{journal}{Annals of Physics} \bibinfo{volume}{458}, \bibinfo{pages}{169453}.
\newblock \URLprefix \url{https://www.sciencedirect.com/science/article/pii/S0003491623002555}, \DOIprefix\doi{https://doi.org/10.1016/j.aop.2023.169453}.
\bibitem[{Homan et~al.(2009)Homan, Lister, Aller, Aller and Wardle}]{Homan_2009}
\bibinfo{author}{Homan, D.C.}, \bibinfo{author}{Lister, M.L.}, \bibinfo{author}{Aller, H.D.}, \bibinfo{author}{Aller, M.F.}, \bibinfo{author}{Wardle, J.F.C.}, \bibinfo{year}{2009}.
\newblock \bibinfo{title}{Full polarization spectra of 3c 279}.
\newblock \bibinfo{journal}{The Astrophysical Journal} \bibinfo{volume}{696}, \bibinfo{pages}{328}.
\newblock \URLprefix \url{https://dx.doi.org/10.1088/0004-637X/696/1/328}, \DOIprefix\doi{10.1088/0004-637X/696/1/328}.
\bibitem[{Kittel(2004)}]{Kittel}
\bibinfo{author}{Kittel, C.}, \bibinfo{year}{2004}.
\newblock \bibinfo{title}{Introduction to Solid State Physics}.
\newblock \bibinfo{edition}{8} ed., \bibinfo{publisher}{Wiley}.
\bibitem[{Kumar and Zhang(2015)}]{KUMAR20151}
\bibinfo{author}{Kumar, P.}, \bibinfo{author}{Zhang, B.}, \bibinfo{year}{2015}.
\newblock \bibinfo{title}{The physics of gamma-ray bursts \& relativistic jets}.
\newblock \bibinfo{journal}{Physics Reports} \bibinfo{volume}{561}, \bibinfo{pages}{1--109}.
\newblock \URLprefix \url{https://www.sciencedirect.com/science/article/pii/S0370157314003846}, \DOIprefix\doi{https://doi.org/10.1016/j.physrep.2014.09.008}. \bibinfo{note}{the physics of gamma-ray bursts \& relativistic jets}.
\bibitem[{Kunz(1915)}]{Kunz}
\bibinfo{author}{Kunz, J.}, \bibinfo{year}{1915}.
\newblock \bibinfo{title}{On the present theory of magnetism}.
\newblock \bibinfo{journal}{Phys. Rev.} \bibinfo{volume}{6}, \bibinfo{pages}{113--125}.
\newblock \URLprefix \url{https://link.aps.org/doi/10.1103/PhysRev.6.113}, \DOIprefix\doi{10.1103/PhysRev.6.113}.
\bibitem[{Landau and Lifshitz(1960)}]{landau1960electrodynamics}
\bibinfo{author}{Landau, L.D.}, \bibinfo{author}{Lifshitz, E.M.}, \bibinfo{year}{1960}.
\newblock \bibinfo{title}{Electrodynamics of continous media}.
\newblock \bibinfo{publisher}{Pergamon}.
\bibitem[{Landau and Lifshitz(1980)}]{Landau}
\bibinfo{author}{Landau, L.D.}, \bibinfo{author}{Lifshitz, E.M.}, \bibinfo{year}{1980}.
\newblock \bibinfo{title}{{Statistical Physics}}.
\newblock \bibinfo{publisher}{Butterworth-Heinemann}, \bibinfo{address}{Oxford}.
\bibitem[{Lee et~al.(2024)Lee, Rodriguez-Lopez and Woods}]{Woods2024}
\bibinfo{author}{Lee, D.N.}, \bibinfo{author}{Rodriguez-Lopez, P.}, \bibinfo{author}{Woods, L.M.}, \bibinfo{year}{2024}.
\newblock \bibinfo{title}{{Phonon-assisted Casimir interactions between piezoelectric materials}}.
\newblock \bibinfo{journal}{Commun. Mater.} \bibinfo{volume}{5}, \bibinfo{pages}{260}.
\newblock \URLprefix \url{https://doi.org/10.1038/s43246-024-00701-2}, \DOIprefix\doi{10.1038/s43246-024-00701-2}.
\bibitem[{Lifshitz(1956)}]{lifshitz}
\bibinfo{author}{Lifshitz, E.M.}, \bibinfo{year}{1956}.
\newblock \bibinfo{title}{The theory of molecular attractive forces between solids}.
\newblock \bibinfo{journal}{Sov. Phys. JETP} \bibinfo{volume}{2}, \bibinfo{pages}{73}.
\bibitem[{Maia~Neto et~al.(2008)Maia~Neto, Lambrecht and Reynaud}]{NetoPhysRevA}
\bibinfo{author}{Maia~Neto, P.A.}, \bibinfo{author}{Lambrecht, A.}, \bibinfo{author}{Reynaud, S.}, \bibinfo{year}{2008}.
\newblock \bibinfo{title}{Casimir energy between a plane and a sphere in electromagnetic vacuum}.
\newblock \bibinfo{journal}{Phys. Rev. A} \bibinfo{volume}{78}, \bibinfo{pages}{012115}.
\newblock \URLprefix \url{https://link.aps.org/doi/10.1103/PhysRevA.78.012115}, \DOIprefix\doi{10.1103/PhysRevA.78.012115}.
\bibitem[{Ninham and Bostr\"om(2003)}]{PhysRevA.67.030701}
\bibinfo{author}{Ninham, B.W.}, \bibinfo{author}{Bostr\"om, M.}, \bibinfo{year}{2003}.
\newblock \bibinfo{title}{Screened {C}asimir force at finite temperatures: A possible role in nuclear interactions}.
\newblock \bibinfo{journal}{Phys. Rev. A} \bibinfo{volume}{67}, \bibinfo{pages}{030701}.
\newblock \URLprefix \url{https://link.aps.org/doi/10.1103/PhysRevA.67.030701}, \DOIprefix\doi{10.1103/PhysRevA.67.030701}.
\bibitem[{Ninham et~al.(2014)Ninham, Bostr\"om, Persson, Brevik, Buhmann and Sernelius}]{EPJDNinham2014}
\bibinfo{author}{Ninham, B.W.}, \bibinfo{author}{Bostr\"om, M.}, \bibinfo{author}{Persson, C.}, \bibinfo{author}{Brevik, I.}, \bibinfo{author}{Buhmann, S.Y.}, \bibinfo{author}{Sernelius, B.E.}, \bibinfo{year}{2014}.
\newblock \bibinfo{title}{{C}asimir forces in a plasma: Possible connections to {Y}ukawa potentials}.
\newblock \bibinfo{journal}{Eur. Phys. J. D} \bibinfo{volume}{68}, \bibinfo{pages}{328}.
\bibitem[{Ninham et~al.(2022)Ninham, Brevik and Bostr\"om}]{Ninham_Brevik_Bostrom_2022}
\bibinfo{author}{Ninham, B.W.}, \bibinfo{author}{Brevik, I.}, \bibinfo{author}{Bostr\"om, M.}, \bibinfo{year}{2022}.
\newblock \bibinfo{title}{Equivalence of electromagnetic fluctuation and nuclear ({Y}ukawa) forces: the $\pi_0$ meson, its mass and lifetime}.
\newblock \bibinfo{journal}{Substantia} \bibinfo{volume}{7}, \bibinfo{pages}{7--14}.
\newblock \URLprefix \url{https://riviste.fupress.net/index.php/subs/article/view/1807}, \DOIprefix\doi{10.36253/Substantia-1807}.
\bibitem[{Ninham and Daicic(1998)}]{NinhamDaicicPhysRevA.57.1870}
\bibinfo{author}{Ninham, B.W.}, \bibinfo{author}{Daicic, J.}, \bibinfo{year}{1998}.
\newblock \bibinfo{title}{Lifshitz theory of {C}asimir forces at finite temperature}.
\newblock \bibinfo{journal}{Phys. Rev. A} \bibinfo{volume}{57}, \bibinfo{pages}{1870--1880}.
\newblock \URLprefix \url{https://link.aps.org/doi/10.1103/PhysRevA.57.1870}, \DOIprefix\doi{10.1103/PhysRevA.57.1870}.
\bibitem[{Ninham et~al.(1966)Ninham, Powell and Swanson}]{NinhamPhysRev.145.209}
\bibinfo{author}{Ninham, B.W.}, \bibinfo{author}{Powell, C.J.}, \bibinfo{author}{Swanson, N.}, \bibinfo{year}{1966}.
\newblock \bibinfo{title}{Plasmon damping in metals}.
\newblock \bibinfo{journal}{Phys. Rev.} \bibinfo{volume}{145}, \bibinfo{pages}{209--217}.
\newblock \URLprefix \url{https://link.aps.org/doi/10.1103/PhysRev.145.209}, \DOIprefix\doi{10.1103/PhysRev.145.209}.
\bibitem[{Parsegian and Ninham(1969)}]{ParNin1969}
\bibinfo{author}{Parsegian, V.}, \bibinfo{author}{Ninham, B.}, \bibinfo{year}{1969}.
\newblock \bibinfo{title}{Application of the lifshitz theory to the calculation of van der waals forces across thin lipid films}.
\newblock \bibinfo{journal}{Nature} \bibinfo{volume}{224}, \bibinfo{pages}{1197--1198}.
\newblock \URLprefix \url{https://doi.org/10.1038/2241197a0}, \DOIprefix\doi{10.1038/2241197a0}.
\bibitem[{Rodriguez-Lopez et~al.(2020)Rodriguez-Lopez, Popescu, Fialkovsky, Khusnutdinov and Woods}]{Woods2020}
\bibinfo{author}{Rodriguez-Lopez, P.}, \bibinfo{author}{Popescu, A.}, \bibinfo{author}{Fialkovsky, I.}, \bibinfo{author}{Khusnutdinov, N.}, \bibinfo{author}{Woods, L.M.}, \bibinfo{year}{2020}.
\newblock \bibinfo{title}{{Signatures of complex optical response in Casimir interactions of type I and II Weyl semimetals}}.
\newblock \bibinfo{journal}{Commun. Mater.} \bibinfo{volume}{1}, \bibinfo{pages}{14}.
\newblock \URLprefix \url{https://doi.org/10.1038/s43246-020-0015-4}, \DOIprefix\doi{10.1038/s43246-020-0015-4}.
\bibitem[{Schoger et~al.(2022)Schoger, Spreng, Ingold, Lambrecht, Maia~Neto and Reynaud}]{twosphere_Schoger_IntJModPhysA2022}
\bibinfo{author}{Schoger, T.}, \bibinfo{author}{Spreng, B.}, \bibinfo{author}{Ingold, G.L.}, \bibinfo{author}{Lambrecht, A.}, \bibinfo{author}{Maia~Neto, P.A.}, \bibinfo{author}{Reynaud, S.}, \bibinfo{year}{2022}.
\newblock \bibinfo{title}{Universal {C}asimir interactions in the sphere–sphere geometry}.
\newblock \bibinfo{journal}{Int. J. Mod. Phys. A} \bibinfo{volume}{37}, \bibinfo{pages}{2241005}.
\newblock \URLprefix \url{https://doi.org/10.1142/S0217751X22410056}, \DOIprefix\doi{10.1142/S0217751X22410056}, \href{http://arxiv.org/abs/https://doi.org/10.1142/S0217751X22410056}{{\tt arXiv:https://doi.org/10.1142/S0217751X22410056}}.
\bibitem[{Sernelius(2018)}]{Ser2018}
\bibinfo{author}{Sernelius, B.E.}, \bibinfo{year}{2018}.
\newblock \bibinfo{title}{{Fundamentals of van der Waals and Casimir Interactions}}.
\newblock Springer Series on Atomic, Optical, and Plasma Physics, \bibinfo{publisher}{Springer International Publishing}.
\newblock \URLprefix \url{https://www.springer.com/us/book/9783319998305}, \DOIprefix\doi{10.1007/978-3-319-99831-2}.
\bibitem[{Shelden et~al.(2023)Shelden, Spreng and Munday}]{shelden2023enhanced}
\bibinfo{author}{Shelden, C.}, \bibinfo{author}{Spreng, B.}, \bibinfo{author}{Munday, J.N.}, \bibinfo{year}{2023}.
\newblock \bibinfo{title}{Enhanced repulsive {C}asimir forces between gold and thin magnetodielectric plates}.
\newblock \bibinfo{journal}{Physical Review A} \bibinfo{volume}{108}, \bibinfo{pages}{032817}.
\bibitem[{Shelden et~al.(2024)Shelden, Spreng and Munday}]{MundayReview2024}
\bibinfo{author}{Shelden, C.}, \bibinfo{author}{Spreng, B.}, \bibinfo{author}{Munday, J.N.}, \bibinfo{year}{2024}.
\newblock \bibinfo{title}{Opportunities and challenges involving repulsive {C}asimir forces in nanotechnology}.
\newblock \bibinfo{journal}{Appl. Phys. Rev.} \bibinfo{volume}{11}, \bibinfo{pages}{041325}.
\newblock \URLprefix \url{https://doi.org/10.1063/5.0218274}, \DOIprefix\doi{10.1063/5.0218274}.
\bibitem[{Spreng et~al.(2022)Spreng, Gong and Munday}]{MundayTorque2022}
\bibinfo{author}{Spreng, B.}, \bibinfo{author}{Gong, T.}, \bibinfo{author}{Munday, J.N.}, \bibinfo{year}{2022}.
\newblock \bibinfo{title}{Recent developments on the {C}asimir torque}.
\newblock \bibinfo{journal}{Int. J. Mod. Phys. A} \bibinfo{volume}{37}, \bibinfo{pages}{2241011}.
\newblock \URLprefix \url{https://doi.org/10.1142/S0217751X22410111}, \DOIprefix\doi{10.1142/S0217751X22410111}.
\bibitem[{Stoneking et~al.(2020)Stoneking, Pedersen, Helander, Chen, Hergenhahn, Stenson, Fiksel, von~der Linden, Saitoh, Surko and et~al.}]{Stoneking_Pedersen_Helander}
\bibinfo{author}{Stoneking, M.R.}, \bibinfo{author}{Pedersen, T.S.}, \bibinfo{author}{Helander, P.}, \bibinfo{author}{Chen, H.}, \bibinfo{author}{Hergenhahn, U.}, \bibinfo{author}{Stenson, E.V.}, \bibinfo{author}{Fiksel, G.}, \bibinfo{author}{von~der Linden, J.}, \bibinfo{author}{Saitoh, H.}, \bibinfo{author}{Surko, C.M.}, \bibinfo{author}{et~al.}, \bibinfo{year}{2020}.
\newblock \bibinfo{title}{A new frontier in laboratory physics: magnetized electron–positron plasmas}.
\newblock \bibinfo{journal}{Journal of Plasma Physics} \bibinfo{volume}{86}, \bibinfo{pages}{155860601}.
\newblock \DOIprefix\doi{10.1017/S0022377820001385}.
\bibitem[{{The CLAS Collaboration}(2019)}]{NuclearBinding_CLAS2019}
\bibinfo{author}{{The CLAS Collaboration}}, \bibinfo{year}{2019}.
\newblock \bibinfo{title}{Modified structure of protons and neutrons in correlated pairs}.
\newblock \bibinfo{journal}{Nature} \bibinfo{volume}{566}, \bibinfo{pages}{354–358}.
\newblock \DOIprefix\doi{10.1038/s41586-019-0925-9}.
\bibitem[{Van~Vleck(1932)}]{van1932theory}
\bibinfo{author}{Van~Vleck, J.H.}, \bibinfo{year}{1932}.
\newblock \bibinfo{title}{The theory of electric and magnetic susceptibilities}.
\newblock \bibinfo{publisher}{Oxford University Press}.
\bibitem[{Wick(1938)}]{Wick}
\bibinfo{author}{Wick, G.}, \bibinfo{year}{1938}.
\newblock \bibinfo{title}{{Range of Nuclear Forces in Yukawa's Theory}}.
\newblock \bibinfo{journal}{Nature} \bibinfo{volume}{142}, \bibinfo{pages}{993--994}.
\newblock \URLprefix \url{https://doi.org/10.1038/142993b0}.
\bibitem[{Woods et~al.(2016)Woods, Dalvit, Tkatchenko, Rodriguez-Lopez, Rodriguez and Podgornik}]{WoodsRevModPhys.88.045003}
\bibinfo{author}{Woods, L.M.}, \bibinfo{author}{Dalvit, D.A.R.}, \bibinfo{author}{Tkatchenko, A.}, \bibinfo{author}{Rodriguez-Lopez, P.}, \bibinfo{author}{Rodriguez, A.W.}, \bibinfo{author}{Podgornik, R.}, \bibinfo{year}{2016}.
\newblock \bibinfo{title}{{Materials perspective on Casimir and van der Waals interactions}}.
\newblock \bibinfo{journal}{Rev. Mod. Phys.} \bibinfo{volume}{88}, \bibinfo{pages}{045003}.
\newblock \URLprefix \url{https://link.aps.org/doi/10.1103/RevModPhys.88.045003}, \DOIprefix\doi{10.1103/RevModPhys.88.045003}.
\bibitem[{Yndurain(2012)}]{Yndurain}
\bibinfo{author}{Yndurain, F.J.}, \bibinfo{year}{2012}.
\newblock \bibinfo{title}{Relativistic Quantum Mechanics and Introduction to Field Theory}.
\newblock \bibinfo{publisher}{Springer Berlin Heidelberg}.
\bibitem[{Yukawa(1935)}]{Yukawa1935}
\bibinfo{author}{Yukawa, H.}, \bibinfo{year}{1935}.
\newblock \bibinfo{title}{On the interaction of elementary particles. i}.
\newblock \bibinfo{journal}{Proceedings of the Physico-Mathematical Society of Japan. 3rd Series} \bibinfo{volume}{17}, \bibinfo{pages}{48--57}.
\newblock \DOIprefix\doi{10.11429/ppmsj1919.17.0_48}.
\bibitem[{Yukawa(1955)}]{Yukawa1955}
\bibinfo{author}{Yukawa, H.}, \bibinfo{year}{1955}.
\newblock \bibinfo{title}{{On the Interaction of Elementary Particles. I}}.
\newblock \bibinfo{journal}{Progress of Theoretical Physics Supplement} \bibinfo{volume}{1}, \bibinfo{pages}{1--10}.
\newblock \URLprefix \url{https://doi.org/10.1143/PTPS.1.1}, \DOIprefix\doi{10.1143/PTPS.1.1}.
\bibitem[{Zhang et~al.(2024)Zhang, Zhang, Wang, Wang, Liu, Li, Zhang, Fan and Zeng}]{ZhangNature2024}
\bibinfo{author}{Zhang, Y.}, \bibinfo{author}{Zhang, H.}, \bibinfo{author}{Wang, X.}, \bibinfo{author}{Wang, Y.}, \bibinfo{author}{Liu, Y.}, \bibinfo{author}{Li, S.}, \bibinfo{author}{Zhang, T.}, \bibinfo{author}{Fan, C.}, \bibinfo{author}{Zeng, C.}, \bibinfo{year}{2024}.
\newblock \bibinfo{title}{Magnetic-field tuning of the {C}asimir force}.
\newblock \bibinfo{journal}{Nat. Phys.} \bibinfo{volume}{20}, \bibinfo{pages}{1282--1287}.
\newblock \DOIprefix\doi{10.1038/s41567-024-02521-0}.
\bibitem[{van Zwol and Palasantzas(2010)}]{Zwol1}
\bibinfo{author}{van Zwol, P.J.}, \bibinfo{author}{Palasantzas, G.}, \bibinfo{year}{2010}.
\newblock \bibinfo{title}{{Repulsive Casimir forces between solid materials with high-refractive-index intervening liquids}}.
\newblock \bibinfo{journal}{Phys. Rev. A} \bibinfo{volume}{81}, \bibinfo{pages}{062502}.
\newblock \URLprefix \url{https://link.aps.org/doi/10.1103/PhysRevA.81.062502}, \DOIprefix\doi{10.1103/PhysRevA.81.062502}.

\end{thebibliography}

\end{document}